\documentclass[12pt]{iopart}

\usepackage{graphicx}
\usepackage{psfrag}
\usepackage{amssymb}
\usepackage{iopams}

\begin{document}

\title[Theoretical Analysis of Astronomical Phased Arrays]{Theoretical 
Analysis of Astronomical Phased Arrays}

\author{Stafford Withington, George Saklatavala, and Michael P. Hobson}

\address{Cavendish Laboratory, JJ Thompson Avenue, Cambridge CB3
OHE, UK} \ead{stafford@mrao.cam.ac.uk}

\begin{abstract} Low-noise phased arrays are essential for the next 
generation of microwave and submillimetre wave astronomy. We analyze 
their behaviour from a functional perspective, and show 
that their operation is intimately related to the mathematical theory 
of frames. No assumptions are made about the orthogonality or linear
independence of the synthesised beams. Frame theory allows an unambiguous
assessment of whether the outputs of an array can be used to
observe a field or brightness distribution within a given class.
Image reconstruction is carried out using dual beams. We identify the 
natural modes of phased arrays, and carry out an analysis of noise. The 
scheme allows the expectation values, the mean-square fluctuations, and 
the correlations between fluctuations at the output ports of a phased 
array to be determined for a source in any state of spatial coherence. 
Both classical and photon-counting statistics are included. Our model 
is conceptually powerful, and suggests many simulation and image recovery 
techniques.

\end{abstract}


\maketitle

\pagebreak

\section{Introduction}

Recently, there has been a surge of interest in
developing phased arrays for radio astronomy. Projects include the
Square Kilometer Array (SKA), the Low Frequency Array (LOFAR), the
Electronic Multibeam Radio Astronomy Concept (EMBRACE), the Karoo
Array Telescope (KAT) \cite{A,B,C,D}, and a number of instruments
for enhancing the performance of single-dish telescopes. Most
of the current projects are aimed constructing phased arrays for
microwave astronomy, but as technological capability improves, 
phased arrays will also be constructed for far infrared and 
submillimetre wave astronomy \cite{E,F}.

Two types of phased array are of interest: (i) imaging phased
arrays, where an array of passive antennas, or coherent receivers, 
is connected to a beam-forming network such that synthesised beams 
can be created and swept across the sky; (ii) interferometric phased 
arrays, where the individual antennas of an aperture synthesis
interferometer are equipped with phased arrays such that fringes
are formed within the synthesised beams. In this way it is
possible to extend the field of view, to observe completely
different regions of the sky simultaneously, to steer the field of
view electronically, and to observe spatial frequencies that are
not available because the baselines of an interferometer cannot be
packed more tightly than the diameters of the individual telescopes.

It is important to appreciate that the synthesised beams of a
phased array need not, mathematically speaking, be orthogonal and
may even be linearly dependent. Non-orthogonality may be built
into a system intentionally as a way of increasing the fidelity
with which an image can be reconstructed, or it may arise
inadvertently as a consequence of RF coupling and post-processing 
cross-talk. In some situations, say in the case of interacting 
planar antennas, it may not even be clear how to distinguish one 
antenna from another, even before the beam-forming network has 
been connected.

Despite the considerable importance of phased arrays for astronomy, 
there is still a lack of information about the principles of operation
when the synthesised beams are non-orthogonal, or even 
linearly dependent, and when noise is included. In this paper we 
describe the operation of low-noise imaging phased arrays from a 
functional perspective, and show that information throughput and image 
recovery are intimately related to the mathematical theory of frames. We 
also show that it is only necessary to know the synthesised beams in order 
to calculate the average powers, the correlations between the complex 
travelling wave amplitudes, the fluctuations in power, and the 
correlations between the fluctuations in power (the Hanbury Brown-Twiss 
effect) at the output ports of an array.

The paper provides a powerful conceptual framework for understanding 
the operation of phased arrays. 
A key feature of the analysis is that we do not fill the half space 
between the aperture plane and the sky with a complete set of plane 
waves, but limit our attention to the natural optical modes of the 
system. In this way, issues relating to information throughput, 
sensitivity, noise, correlations, and quantum statistics can be dealt 
with in a straightforward manner. Indeed, the quantum statistical 
properties of the incoming radiation field can be taken into 
account, for a source in any state of spatial coherence, and the 
transition from fully bunched to photon-counting statistics included 
as the wavelength of operation moves from the microwave range 
through into the far infrared.

\section{Basic Principles} \label{sec1}

In general an imaging phased array comprises a sequence of optical
components, an array of single-mode receivers, each of which has
a primary beam pattern, and an electrical beam-forming 
network such that each output port corresponds to a
synthesised reception pattern on the sky. The synthesised
reception patterns may be static and designed to give optimum
sampling on a given class of object, or they may be controlled
electrically and swept across the field of view. In the case of
radio astronomy, the optical system would be a telescope, the
single-mode receivers would be horns or planar antennas coupled to
HEMT amplifiers or SIS mixers, and the beam-forming network would
be a system of microwave or digital electronics.

The following analysis is based on a generic system comprising
an array of $M$ horns and a beam-forming network having $P$ output 
ports. Each of the $P$ ports is associated with a synthesised 
reception pattern. For convenience, ${\cal A}$ denotes the input 
reference surface, which is a far-field region of the sky, 
${\cal B}$ the output ports of the horns, and ${\cal C}$ the output 
ports of the beam-forming network.

When a fully coherent field is incident on the system, a set of
travelling waves appears at ${\cal B}$, and we shall denote their
complex amplitudes by $\{ y_{m}: m\in 1,\cdots,M\}$; also, a set
of travelling waves appears at ${\cal C}$, and we shall denote
their complex amplitudes by $\{ z_{p}: p\in 1,\cdots,P\}$. Because
$M$ and $P$ are finite, the complex amplitudes can be assembled
into column vectors ${\bf y} \in {\mathbb C}^{M}$ and ${\bf z} \in
{\mathbb C}^{P}$. If ${\bf E}(\hat{\boldsymbol \Omega})$ 
is the plane-wave spectrum of the incident electric field, 
and $\hat{\boldsymbol \Omega}$ a unit radial vector in the aperture 
plane pointing towards the sky, then it can be shown that because 
the synthesised reception patterns and incoming field are 
square-integrable functions, which can be represented by
vectors in the Hilbert space of square integral functions over $\cal A$, 
the complex amplitude of the travelling wave at port $p$ can be written
\begin{equation}
\label{A_1}
z_{p} = \int_{\cal A} {\bf t}_{p}^{\ast} ( \hat{ \boldsymbol \Omega }) 
\cdot {\bf E}(\hat{ \boldsymbol \Omega }) \, d { \boldsymbol \Omega } 
\mbox{,}
\end{equation}
where ${\bf t}_{p} ( \hat{\boldsymbol \Omega})$ is the complex 
synthesised reception pattern of port $p$. It would be naive to assume, 
however, that if the array is illuminated by a field having the form 
${\bf E}(\hat{ \boldsymbol \Omega }) = {\bf t}_{p} (\hat{\boldsymbol 
\Omega})$, a travelling wave only appears at $p$.

It is instructive to derive a relationship between the synthesised beams
and the beams of the primary antennas. If the beam patterns of the primary 
antennas are denoted by ${\bf h}_{m} (\hat{\boldsymbol \Omega})$ then the 
outputs of the antennas, $y_{m}$, are given by
\begin{equation} \label{A_2} y_{m} = \int_{\cal A} {\bf
h}^{\ast}_{m} (\hat{\boldsymbol \Omega}) \cdot {\bf E}(\hat{\boldsymbol 
\Omega}) \, d {\boldsymbol  \Omega} \mbox{,}
\end{equation}
but the beam-forming network can be described by scattering matrix
${\boldsymbol \Phi}$, having elements $\phi_{pm}$, where
\begin{equation} \label{A_3} z_{p} = \sum_{m} \phi_{pm} y_{m}
\mbox{,}
\end{equation}
and therefore substituting (\ref{A_2}) in (\ref{A_3}),
\begin{equation}
\label{A_4} z_{p} = \int_{\cal A}  \sum_{m} \phi_{pm} {\bf
h}^{\ast}_{m} (\hat{\boldsymbol \Omega}) \cdot 
{\bf E}(\hat{\boldsymbol \Omega}) \, d {\boldsymbol \Omega} \mbox{,}
\end{equation}
which can be cast into the form of (\ref{A_1}) by defining
\begin{equation}
\label{A_5} {\bf t}_{p} (\hat{\boldsymbol \Omega}) = \sum_{m}
\phi_{pm}^{\ast}  {\bf h}_{m} (\hat{\boldsymbol \Omega}) \mbox{.}
\end{equation}
As expected, the synthesised reception patterns are
weighted linear combinations of the primary antenna patterns.

Even when the primary beams are orthogonal, and identical, the 
synthesised beams do not have to be orthogonal, because
\begin{equation}
\label{A_6} \int_{\cal A}  {\bf t}_{p}^{\ast} 
(\hat{\boldsymbol \Omega}) \cdot {\bf t}_{p'} 
(\hat{\boldsymbol \Omega})  \, d {\boldsymbol \Omega} = 
k \sum_{m} \phi_{pm} \phi_{p'm}^{\ast} \mbox{,}
\end{equation}
where (\ref{A_5}) has been used, $k$ is a constant that derives from
the inner products of the primary antenna patterns, 
\begin{equation}
\int_{\cal A} {\bf h}_{m}^{\ast} (\hat{\boldsymbol \Omega})
\cdot {\bf h}_{m'} (\hat{\boldsymbol \Omega}) \, d {\boldsymbol \Omega} 
= k \, \delta_{mm'}
\mbox{,}
\end{equation}
and there is no restriction on the orthogonality of the $\phi_{pm}$. 
(\ref{A_6}) can be written in the form of a matrix equation:
\begin{equation}
\label{A_7} \int_{\cal A} {\bf t}_{p}^{\ast} (\hat{\boldsymbol \Omega})
\cdot {\bf t}_{p'} (\hat{\boldsymbol \Omega}) \, d {\boldsymbol \Omega} 
= k \left[ {\boldsymbol \Phi}{\boldsymbol \Phi}^{\dagger} 
\right]_{p \, p'} \mbox{.}
\end{equation}
If $M$ of the columns of ${\boldsymbol \Phi}^{\dagger}:{\mathbb C}^{P} 
\rightarrow {\mathbb C}^{M}$ are linearly independent, and $P > M$, then 
the complete set of columns constitutes an over-complete basis, in the sense
that there are $P$ vectors in an $M$-dimensional space. In this case, 
except trivially when certain ports are not connected, the synthesised 
reception patterns are not orthogonal, because 
${\boldsymbol \Phi}{\boldsymbol \Phi}^{\dagger} \neq {\bf D}$, where 
${\bf D}$ is a diagonal matrix of dimension $P$; indeed, the 
synthesised reception patterns are linearly dependent. Contrariwise, the 
columns of ${\boldsymbol \Phi}^{\dagger}$ comprise an under-complete basis 
if $P<M$, and again we may have ${\boldsymbol \Phi}{\boldsymbol 
\Phi}^{\dagger} \neq {\bf D}$. In the case where ${\boldsymbol \Phi}$ is 
unitary, ${\boldsymbol \Phi}{\boldsymbol \Phi}^{\dagger} = {\bf I}_{P}$, 
where ${\bf I}_{P}$ is the identity matrix of dimension $P$,
and the primary beam patterns are orthogonal, the synthesised reception 
patterns are orthogonal. We conclude that, except in the simplest of 
scenarios, the synthesised reception patterns are unlikely to be orthogonal,
\begin{equation}
\label{A_8} \int_{\cal A} {\bf t}_{p}^{\ast} ( 
\hat{\boldsymbol \Omega}) \cdot {\bf t}_{p'} 
( \hat{\boldsymbol \Omega}) \, d {\boldsymbol \Omega} 
\neq k D_{p,p'} 
\mbox{,}
\end{equation}
and may even be linearly dependent. According to (\ref{A_1}) and 
(\ref{A_8}), travelling waves will generally appear at more than 
one port even when the incident field has the form of one of the 
synthesised reception patterns: ${\bf E}(\hat{\bf \Omega}) = 
{\bf t}_{p'} ( \hat{\bf \Omega})$.

(\ref{A_5}) indicates that the synthesised beams are weighted linear
combinations of the primary antenna patterns; or conversely, that
there must be sufficient antennas of the correct form, such that
all of the required synthesised reception patterns, are spanned, 
mathematically speaking, by the primary antenna patterns. 
In the case of real systems, the primary beams ${\bf h}_{m}
(\hat{\bf \Omega}): m \in \{ 1, \cdots, M \}$, which are the
individual beams in the presence of all of the antennas, including
scattering, are likely to be different to one another; for
example, the antennas at the edge of an array will have different
beam patterns to those in the middle. In addition, the primary
beams may have far-out sidelobes, which can couple to warm
radiating objects such as the ground. Ideally, it must be possible
to span all possible synthesised beams, as the source moves across
the sky, including the requirement that the synthesised beams must
be near zero outside of some field of view. If nulling is not 
achieved, high noise temperatures, which may be a strong function of 
${\bf \Phi}$, may result. Low noise temperatures can only be achieved 
by ensuring that the reception patterns of the primary antennas can
only couple to low-noise regions of the sky, or, formally
speaking, that there are many more degrees of freedom in the
system than are needed to simply create the main lobes of
the synthesised beams.

In astronomical applications, phased arrays are used to image
incoherent or, in the case of celestial masers, partially coherent
fields. It is convenient, therefore, to introduce correlation
dyadics. Define the correlation dyadic of the incident field 
according to
\begin{equation}
\label{A_9} \overline{\overline{\bf E}}(\hat{\boldsymbol
\Omega}_{1}, \hat{\boldsymbol \Omega}_{2}) = \langle {\bf E}
(\hat{\boldsymbol \Omega}_{1}) \hat{\bf E}^{\ast}
(\hat{\boldsymbol \Omega}_{2}) \rangle
\mbox{,}
\end{equation}
where ${\bf E}(\hat{\boldsymbol \Omega})$ is
the complex analytic representation of the quasi-monochromatic
electric field, and $\langle \, \, \rangle$ denotes the ensemble
average. The final result can be integrated with respect to
frequency to calculate broadband behaviour, but we do not show 
frequency dependence explicitly. We shall assume throughout 
that the electric-field is normalized to the square root of the 
impedance of free space so that the elements of 
$\overline{\overline{\bf E}}(\hat{\boldsymbol 
\Omega}_{1}, \hat{\boldsymbol \Omega}_{2})$ have the units of
Wm$^{-2}$Sr$^{-2}$Hz$^{-1}$. 
The rank 2 tensor $\overline{\overline{\bf E}}(\hat{\boldsymbol 
\Omega}_{1}, \hat{\boldsymbol \Omega}_{2})$ contains complete 
information about the correlations between all pairs of transverse 
vector field components for any two points on the sky. 
Once the correlation dyadic is known, all classical measures of 
coherence follow. 

The correlation between the travelling wave 
amplitudes at any two output ports is given by
$\langle z_{p} z_{p'}^{\ast} \rangle$, or in matrix form ${\bf Z}
=  \langle {\bf z}{\bf z}^{\dagger} \rangle \in
{\mathbb C}^{P \times P}$. The elements of ${\bf Z}$ can be
found by using (\ref{A_1}) and (\ref{A_9}):
\begin{equation}
\label{A_13} Z_{pp'} = \int_{\cal A} \int_{\cal A} {\bf
t}_{p}^{\ast} (\hat{\boldsymbol \Omega}_{1}) \cdot 
\overline{\overline{\bf E}}(\hat{\boldsymbol \Omega}_{1}, 
\hat{\boldsymbol \Omega}_{2}) \cdot {\bf t}_{p'}
({\boldsymbol \Omega}_{2}) \, d {\boldsymbol \Omega}_{1} \, 
d {\boldsymbol \Omega}_{2} \mbox{,}
\end{equation}
and because the the travelling wave amplitudes are normalized to 
the square root of impedance, the elements of ${\bf Z}$ have the 
units WHz$^{-1}$. According to (\ref{A_13}) the correlations 
at the outputs are merely the matrix elements of the source
coherence tensor with respect to the synthesised beams.

In the case of spatially incoherent, but not necessarily 
unpolarized sources,
\begin{equation}
\label{A_10} \overline{\overline{\bf E}}(\hat{\boldsymbol
\Omega}_{1},\hat{\boldsymbol \Omega}_{2}) = \overline{\overline{\bf
B}}(\hat{\boldsymbol \Omega}_{1}) \delta(\hat{\boldsymbol 
\Omega}_{1}-\hat{\boldsymbol \Omega}_{2}) \mbox{,}
\end{equation}
where $\overline{\overline{\bf B}}(\hat{\boldsymbol \Omega}_{1})$ 
contains information about the polarisation of the source, which 
can be projected onto Pauli-spin matrices or spin-weighted spherical
harmonics. $\overline{\overline{\bf B}}(\hat{\boldsymbol \Omega}_{1})$ 
is a brightness tensor because it has units Wm$^{-2}$Sr$^{-1}$Hz$^{-1}$. 
In the case of incoherent and unpolarised sources,
(\ref{A_10}) becomes
\begin{equation}
\label{A_11} \overline{\overline{\bf E}}(\hat{\bf
\Omega}_{1},\hat{\bf \Omega}_{2}) = B(\hat{\bf \Omega}_{1}) \,
\overline{\overline{\bf I}} \, \delta(\hat{\bf
\Omega}_{1}-\hat{\bf \Omega}_{2}) \mbox{,}
\end{equation}
where $\overline{\overline{\bf I}}$ is the unit dyad, and
$B(\hat{\bf \Omega}_{1})$ the brightness of each polarisation
in the direction of $\hat{\bf \Omega}_{1}$. 

(\ref{A_11}) can be expressed in terms of a brightness
temperature, $T_{b}(\hat{\bf \Omega})$, through
\begin{equation}
\label{A_11b} \overline{\overline{\bf E}}(\hat{\boldsymbol
\Omega}_{1},\hat{\boldsymbol \Omega}_{2}) = 
\frac{h \nu^{3}}{c^{2}} \frac{1}{\exp [ h \nu /
kT_{b} (\hat{\boldsymbol \Omega}_{1})] - 1} \overline{\overline{\bf I}} \,
\delta(\hat{\boldsymbol \Omega}_{1}-\hat{\boldsymbol \Omega}_{2}) \mbox{.}
\end{equation}
(\ref{A_11b}) predicts an infinite correlation dyadic when $\hat{\boldsymbol
\Omega}_{1} = \hat{\boldsymbol \Omega}_{2}$, which occurs because an
infinitely small coherence area requires an infinitely large
number of radiators to be packed into every finite region of the
sky. Nevertheless, (\ref{A_11b}) behaves correctly when integrated
with respect to antenna power patterns, because antenna power
patterns, even in the multimode case, have finite coherence
areas.

Numerous sources can be modelled in this way, for example, a
polarized source can be combined with unpolarized emission from
the atmosphere having brightness temperature $T_{s}$, to give
\begin{equation}
\label{A_12} \overline{\overline{\bf E}}(\hat{\boldsymbol
\Omega}_{1},\hat{\boldsymbol \Omega}_{2}) = \left[ 
\frac{h \nu^{3}}{c^{2}} \frac{1}{\exp [ h
\nu / kT_{s}] - 1} \, \overline{\overline{\bf I}} +
\overline{\overline{\bf B}}(\hat{\boldsymbol \Omega}_{1}) \right] \,
\delta(\hat{\boldsymbol \Omega}_{1}-\hat{\boldsymbol \Omega}_{2}) 
\mbox{.}
\end{equation}
Some care is needed when $T_{s}$ is not uniform because the source
is in the far field of the array whereas the atmosphere is usually in
in the near field; these effects can be taken into account, but we 
shall not do so here.

Now illuminate a phased array with an unpolarised, incoherent source. 
Substituting (\ref{A_11}) in (\ref{A_13}), the elements of the 
coherence matrix become
\begin{equation}
\label{A_14} Z_{pp'} =  \int_{\cal A} B(\hat{\bf \Omega}) \,  {\bf
t}_{p}^{\ast} (\hat{\boldsymbol \Omega}) \cdot {\bf t}_{p'} 
(\hat{\boldsymbol \Omega}) \, d {\boldsymbol \Omega} \mbox{,}
\end{equation}
the diagonal elements of which correspond to coupling the incoming
radiation field to the synthesised power patterns, $K_{p}(\hat{\boldsymbol 
\Omega}) = {\bf t}_{p}^{\ast} (\hat{\boldsymbol \Omega}) \cdot {\bf t}_{p} 
(\hat{\boldsymbol \Omega})$, giving
\begin{equation}
\label{A_14b} Z_{pp} =  \int_{\cal A} B(\hat{\boldsymbol \Omega}) \,  
K_{p}(\hat{\boldsymbol \Omega}) \, d {\boldsymbol \Omega} \mbox{.}
\end{equation}
The functions $K_{p}(\hat{\boldsymbol \Omega})$ are direction-dependent
effective areas, and we could use $K_{p}(\hat{\boldsymbol \Omega})=
A_{p} P_{p}(\hat{\boldsymbol \Omega})$, where $A_{p}$ is the effective 
area of the $p$'th synthesised beam in the most receptive direction, and 
$P_{p}(\hat{\boldsymbol \Omega})$ is the dimensionless power pattern. 
Substituting the Planck formula in (\ref{A_14b}), and assuming
a uniform sky temperature, gives
\begin{eqnarray}
\label{A_14c} Z_{pp} & = &  \int_{\cal A} 
\frac{h \nu^{3}}{c^{2}} \frac{1}{\exp [ h
\nu / kT_{s}] - 1} \, A_{p} P_{p}(\hat{\boldsymbol \Omega})
\, d {\boldsymbol \Omega}  \\ \nonumber
 & = & \frac{ A_{p} \, \Omega_{p}}{\lambda^{2}}
\frac{h \nu}{\exp [ h \nu / kT_{s}] - 1} \\ \nonumber
 & = & \kappa_{p} \frac{h \nu}{\exp [ h \nu / kT_{s}] - 1} 
\mbox{,}
\end{eqnarray}
where $\kappa_{p} = A_{p} \, \Omega_{p} / \lambda^{2}$ is a
coupling factor. For orthogonal primary beams, and a unitary
beam-forming network, $A_{p} \, \Omega_{p} / \lambda^{2}$ can 
be regarded as the number of modes in the beam, which is unity,
and we arrive at the expected spectral power in the travelling 
wave. Generally, however, for a phased array, 
$A_{p} \, \Omega_{p} / \lambda^{2} < 1$. Certainly $\sum_{p} 
A_{p} \, \Omega_{p} / \lambda^{2} \leq M$, because the total 
power cannot exceed the modal throughput of the primary 
antennas.

(\ref{A_14}) shows that because the synthesised reception patterns are 
generally not orthogonal the travelling waves at the output ports are 
correlated, even when the sky brightness is uniform. If the source is not 
uniform, correlations exist even when the synthesised beams are orthogonal, 
which corresponds to the usual case of interferometry. Thus, each port of 
a phased array records the incident flux in the usual manner, but because 
the beams are not necessarily orthogonal, and may even be linearly 
dependent, the travelling waves at the output ports may be 
correlated, and the correlations will be more complicated than in the 
usual case of interferometry. 

\section{Information Throughput and Image Recovery} \label{sec2}

Consider information throughput and image recovery. Represent the primary 
physical quantities as abstract vectors. Because the incoming field, 
${\bf E}(\hat{\boldsymbol \Omega})$, is square integrable over 
the appropriate region of the sky, $\cal A$, it can be
represented by a vector $| {\bf E} \rangle$ in Hilbert space
${\mathbb H}$. Regions having different shapes
and sizes correspond to different Hilbert spaces. 
In the case of realizable phased arrays, where the numbers 
of ports and primary antennas are finite, the measurable quantities 
${\bf y}$ and ${\bf z}$ are finite-dimensional vectors, 
$ {\bf y} \in {\mathbb C}^{M}$ and ${\bf z} \in 
{\mathbb C}^{P}$, but because we wish to draw attention to 
the relationship with frame theory, and to emphasize the role of 
convergence, we use the more general
representations $| {\bf y} \rangle \in {\ell}^{2}$ and 
$|{\bf z} \rangle \in {\ell}^{2}$, where ${\ell}^{2}$ 
is the space of square-summable complex sequences. These 
definitions lead to two operators, one of 
which, $\hat{\bf H}: {\mathbb H} \rightarrow {\ell}^{2}$, maps the
incoming electric field onto the outputs of the horns, and the
other $\hat{\bf \Phi}: {\ell}^{2} \rightarrow {\ell}^{2}$ maps the
outputs of the horns onto the outputs of the beam-forming network.
These individual operators can be combined into a single composite
operator $\hat{\bf T}=\hat{\bf \Phi}\hat{\bf H}:{\mathbb H}
\rightarrow {\ell}^{2}$, which describes the system as a whole.

The operation of phased arrays is intimately related to the
mathematical theory of frames. Suppose that we 
have some general monochromatic field $|{\bf E} \rangle$, and that 
we determine the inner products with respect to a set of basis
vectors  ${\mathbb T} = \{|{\bf t}_{p} \rangle, \, p \in 1,
\cdots, P \}$: $z_{p} = \langle {\bf t}_{p} | {\bf E} \rangle$.
$P$ can extend to infinity, and we do not make any assumptions
about the orthonormality or linear independence of ${\mathbb T}$.
Under what circumstances can the original vector $|{\bf
E}\rangle$, which represents a continuous function, be recovered
unambiguously from a discrete, possibly countable, set of complex 
coefficients, and how can this be achieved? In the context
of phased arrays, we are asking under what circumstances can the 
form of an incident electric field be recovered unambiguously 
from the complex travelling-wave outputs.

Frame theory \cite{G,H,I} proceeds as follows. Evaluate the square 
moduli of the inner products between ${\mathbb
T}$ and any general vector, $|{\bf E}\rangle \in {\mathbb H}$, and 
sum the results. If there are two non-zero constants $A$ and $B$ such 
that $0<A<\infty$ and $0<B<\infty$, and
\begin{equation}
\label{B_1} A \parallel {\bf E} \parallel^{2} \, \leq \,
\parallel \hat{\bf T} | {\bf E} \rangle
\parallel^{2} \, \leq B \, \parallel {\bf E}
\parallel^{2}
\mbox{,}
\end{equation}
which can also be written
\begin{equation}
\label{B_2} A \parallel {\bf E} \parallel^{2} \, \leq \, \sum_{p}
\left| \langle{\bf t}_{p} | {\bf E} \rangle_{\mathbb H}
\right|^{2} \, \leq \, B \parallel {\bf E} \parallel^{2} \mbox{,}
\end{equation}
$\forall \, |{\bf E}\rangle \in {\mathbb H}$, then the basis set
${\mathbb T}$ is called a frame with respect to ${\mathbb H}$.

Notice the strict use of inequalities in the allowable values of
$A$ and $B$. In the case where $A \approx B$, the frame is called
a {\em tight frame} because the inner products for all $|{\bf
E}\rangle \in {\mathbb H}$ lie within some small range, and the
dynamic range needed for inversion is small. When the original
basis is orthonormal, the frame bounds, $A$ and $B$, are equal, as
can be appreciated by inserting $|{\bf E}\rangle = |{\bf
t}_{p'}\rangle$ in (\ref{B_2}). If the frame is normalized, $A$ is
a measure of the redundancy in the frame. If a basis set
constitutes a frame, then it can be shown through (\ref{B_1})
alone that the electric field can be recovered unambiguously from
the inner products, and the more tightly bound the frame, the more
tightly bound the inverse, and the more stable the image recovery
process. The frame condition, (\ref{B_1}), is completely general,
and applies for any continuous function, even though an infinite 
number of possibly linearly dependent basis functions may be used. 
Phased arrays have a finite number of ports,
but can nevertheless span spatially band-limited functions having
finite support, and can therefore form frames with respect to
fields carrying finite information.

Because the synthesised beam
patterns of a phased array may be non-orthogonal, and even
linearly dependent, the recovery of the original field, through an
operator we shall call $\hat{\bf T}^{-1}$, is best implemented
by the introduction of dual vectors, which correspond to dual
beams. The dual vectors $| \widetilde{\bf t}_{p} \rangle$ of any
given frame ${\mathbb T}$, with respect to Hilbert space ${\mathbb
H}$, are given by
\begin{equation}
\label{B_3} | \widetilde{\bf t}_{p} \rangle = \hat{\bf S}^{-1}
|{\bf t}_{p} \rangle \mbox{,}
\end{equation}
where $\hat{\bf S} = \hat{\bf T}^{\dagger} \hat{\bf T}$ is
non-singular, and can therefore be inverted. The dual basis set,
which we shall call $\widetilde{\mathbb T} = \{|\widetilde{\bf
t}_{p} \rangle, \, p \in 1, \cdots, P \}$, has the same degree of
completeness as the original frame, ${\mathbb T}$, and therefore
it too constitutes a frame with respect to ${\mathbb H}$. Indeed,
two representations of any general $|{\bf E}\rangle$ are possible:
\begin{equation}
\label{B_4} |{\bf E}\rangle  =  \sum_{p} \langle {\bf t}_{p} |
{\bf E} \rangle \, |\widetilde{\bf t}_{p} \rangle \hspace{10mm}
|{\bf E}\rangle  =  \sum_{p} \langle \widetilde{\bf t}_{p} | {\bf
E} \rangle \, |{\bf t}_{p} \rangle \mbox{.}
\end{equation}
(\ref{B_4}) shows that if one calculates a set of coefficients by
taking the inner products with a frame, then one inverts the
process by reconstructing the field using the dual vectors. In the
case where the basis vectors are perfectly complete with respect
to ${\mathbb H}$, but not necessarily orthogonal, the basis is
called a Riesz basis, and the basis set ${\mathbb T}$ and dual set
$\widetilde{\mathbb T}$ are biorthogonal: $\langle \widetilde{\bf
t}_{p} | {\bf t}_{p'} \rangle = \delta_{pp'}: \forall \, p,p' \in
1,\cdots,P$.

In the case where the basis vectors do not constitute a frame, but 
an attempt is made to reconstruct the original field vector using the 
duals,
\begin{equation}
\label{B_5} |{\bf E}'\rangle = \sum_{p} \langle {\bf t}_{p} | {\bf
E} \rangle \, |\widetilde{\bf t}_{p} \rangle \mbox{,}
\end{equation}
the reconstructed vector $|{\bf E}'\rangle$ cannot, for all
vectors in ${\mathbb H}$, be the same as the original vector
$|{\bf E}\rangle$. It can be shown, however, that the error vector
$|{\bf E}\rangle-|{\bf E}'\rangle$ is orthogonal to the basis
vectors. Consequently, $|{\bf E}'\rangle$ is the orthogonal
projection of $|{\bf E}\rangle$ onto ${\mathbb S}$, the subspace
spanned by the under complete set of basis vectors. In other words
the solution is as close as possible to the original field vector
to within the degrees of freedom available.

The relevance to phased arrays is clear; one can measure the
complex outputs of a phased array, and if the synthesised
reception patterns constitute a frame with respect to the Hilbert
space defined by the shape, extent, and illumination of the input
reference surface, then the continuous, coherent, incoming field
can be reconstructed completely from the complex travelling wave
amplitudes at the output ports. If the reception patterns do
not constitute a frame, reconstruction leads to the least square
fit that is consistent with the degrees of freedom to which the
phased array is sensitive. If the field of interest has passed
through an optical system, which can only ever transmit a finite
number of modes, then frames can in principle be formed; if
the input surface corresponds to the sky, frames are not 
possible, because the dimensionality of a field, even over a finite
region, is infinite. It is possible, however, to form a frame with 
respect to some given class of object, as will be discussed. It is also 
possible to synthesize more and more beams to tighten a frame, thereby
increasing the stability of the image recovery process. Central to
this model is the notion of dual beams. Every synthesised beam 
has a dual beam, and the concept of dual beams is central to 
understanding the optical physics of phased arrays, and the image 
recovery process.

To this point it has been assumed that the incoming field is fully
coherent, but in the case of astronomical phased arrays the
incoming field is usually incoherent, and sometimes only the
powers are measured at $\cal C$. According to (\ref{A_14}), the
self and cross correlations at the output ports of a phased array,
for an incoherent, unpolarized source are given by
\begin{equation}
\label{B_6} Z_{pp'} =  \int_{\cal A} B(\hat{\boldsymbol \Omega}) \,
S_{pp'}^{\ast} (\hat{\boldsymbol \Omega}) \, d {\boldsymbol \Omega} 
\mbox{,}
\end{equation}
where
\begin{equation}
\label{B_7} S_{pp'}^{\ast} (\hat{\boldsymbol \Omega}) = 
{\bf t}_{p}^{\ast} (\hat{\boldsymbol \Omega}) \cdot 
{\bf t}_{p'} (\hat{\boldsymbol \Omega}) \mbox{,}
\end{equation}
but according to (\ref{B_6}) the elements of the correlation
matrix are merely the inner products of $B(\hat{\boldsymbol
\Omega})$ with respect to $S_{pp'} (\hat{\boldsymbol \Omega}) 
\, \, \forall \, \, p,p' \in \{ 1, \cdots, P \}$. Hence 
once again, frame theory can be used to determine the degree to 
which $ B(\hat{\boldsymbol \Omega})$ can be recovered from the 
correlations between the travelling waves at $\cal C$. The case 
were only the powers are measured, $Z_{pp} \, \, \forall \, \, 
p \in \{ 1, \cdots, P \}$, is a special case of (\ref{B_6}), and 
corresponds to determining the degree to which the synthesised 
power patterns span the source brightness distributions of interest. 
It seems, therefore, that frames can be defined for either fully 
coherent fields with correlation measurement, for incoherent fields 
with correlation measurement, or for incoherent fields with only power 
measurement. We shall call these possibilities {\em field frames}, 
{\em interferometric frames}, and {\em intensity frames} respectively. 
Note that the term {\em interferometric} does not necessarily imply
that an aperture synthesis interferometer is being used, but
merely that the correlations between the travelling waves at the outputs
are measured. Each of these frames can be associated with dual beams of 
some kind. In fact, it is straightforward to formulate a more general 
theory, and ask whether a set of synthesised beams forms a frame with 
respect to the full coherence tensor field of the incoming radiation. 
This more general theory has interesting applications, but we shall not 
described it here.

Suppose that the goal is to reconstruct the intensity distribution
of an astronomical source, and one needs to know whether the basis
$S_{pp'} (\hat{\boldsymbol \Omega}) \, \, \forall \, \, p,p' \in \{ 1,
\cdots, P \}$ forms an interferometric frame. There is a problem,
however, because in assuming that the source is spatially
incoherent, we have tacitly assumed that the brightness
distribution is a member of an infinite dimensional space. To
answer the question correctly, we need to ask whether the phased
array is suitable for recovering brightness from the vector space
of the brightness distributions of interest. One approach is to
describe the range of possible brightness distributions as a finite,
weighted linear combination of basis functions, $\psi_{n}
(\hat{\boldsymbol \Omega}): n \in \{ 1,\cdots,N \}$. These functions
could, for example, be radial basis functions, or delta functions 
at certain sample points. The functions need not correspond to a 
single region, but could correspond to different regions of the 
sky that need to be imaged simultaneously.

The intensity on the sky, according to our chosen class, can be
written
\begin{equation}
\label{B_8} B(\hat{\boldsymbol \Omega}) = \sum_{n} a_{n} \, \psi_{n}
(\hat{\boldsymbol \Omega}) \mbox{,}
\end{equation}
and therefore, according to (\ref{B_6}), the correlations between
the travelling waves at the output ports become
\begin{equation}
\label{B_9} Z_{pp'} = \sum_{n} a_{n} \, F_{pp',n} \mbox{,}
\end{equation}
where
\begin{equation} \label{B_10} F_{pp',n} =  \int_{\cal A}
S_{pp'}^{\ast} (\hat{\boldsymbol \Omega}) \psi_{n} (\hat{\boldsymbol 
\Omega}) \, d {\boldsymbol \Omega} \mbox{.}
\end{equation}

Defining the matrix ${\bf F}$ whose elements are $F_{q,n}$, where
$pp'$ is now indexed by the single integer $q$, the frame
condition (\ref{B_1}) becomes
\begin{equation}
\label{B_11} A \leq {\bf a}^{\dagger} {\bf F}^{\dagger} {\bf F}
{\bf a} \leq B \hspace{5mm} \forall \, {\bf a} \in {\mathbb C}^{N}
\mbox{.}
\end{equation}

(\ref{B_11}) has the Hermitian form, and it is well known that the 
stationary values of the Hermitian form occur when the unit 
vector ${\bf a}$ points in the same directions as the eigenvectors 
of ${\bf F}^{\dagger}{\bf F}$, with the stationary values being 
the corresponding eigenvalues. $A$ and $B$, and hence the tightness 
of the frame, can be determined by finding the largest and smallest
eigenvalues of the Hermitian operator ${\bf F}^{\dagger}{\bf F}$,
which can be found analytically or numerically. Although, the operator 
${\bf F}^{\dagger} {\bf F}$ maps a finite dimensional space onto itself, 
the mapping passes through a space having infinite dimensions and 
therefore the integrals in (\ref{B_10}) should be evaluated analytically 
if at all possible. The operator ${\bf F}^{\dagger} {\bf F}$ simply maps
the intensity distribution coefficients of the source onto the
measured quantities at the output ports of the array and then back
again onto the coefficients. If the basis functions
$S_{pp'} (\hat{\boldsymbol \Omega})$ do not span all possible intensity
distributions, information is lost when an observation is made,
and it is not possible to recover complete information about the
source.

In the case where the basis functions correspond to sample points 
$\hat{\boldsymbol \Omega}_{n}$, we have $\psi_{n}(\hat{\boldsymbol 
\Omega}) = \delta(\hat{\boldsymbol \Omega}-{\boldsymbol \Omega}_{n})$ 
and $F_{pp',n} = S_{pp'}^{\ast} (\hat{\boldsymbol \Omega}_{n})$. 
Although, in practice, a frame cannot be formed that reproduces the sky 
brightness at every one of an infinite number of points, a 
frame can be formed with respect to a finite number of sample points.
Also, shapelets\cite{J,K,L}, which are essentially Gaussian Hermite and 
Gaussian Laguerre polynomials, are now being used to parameterize the 
brightness distributions of astronomical sources, such as galaxies. It 
would therefore be possible to check whether an imaging phased 
array forms a frame with respect to a set of shapelets, and to recover 
the shapelet coefficients directly from observations.

\section{Natural Modes of Phased Arrays}

Superficially speaking, the natural modes of a phased array 
are those orthogonal field distributions that can pass through the 
horns and beam-forming network with high efficiency. When the 
synthesised beams are linearly dependent, or even if they are simply 
non-orthogonal, the forms of the natural modes are not directly known. 
The natural modes are, however, central to understanding noise and 
correlations, to understanding quantum statistical behaviour, which 
is important at sub-millimetre wavelengths, and to understanding the 
operation of interferometric phased arrays.

Suppose that a telescope is equipped with an imaging phased array.
As described previously, the process of projecting a
coherent field onto the complex travelling wave amplitudes at the
output ports is described by the mapping ${\bf T}: |{\bf x}\rangle
\longmapsto |{\bf z}\rangle$ as ${\bf T}:{\mathbb H} \rightarrow
{\ell}^{2}$. The phased array acts as a linear operator between
two Hilbert spaces: the space of square integrable functions over
the input reference surface, and the space of square summable
complex sequences. Because the operator maps between two different
Hilbert spaces, it does not make sense to search for the
eigenfunctions of such an operator in an attempt to find the
natural modes.

For any real system, it is certainly known that every incoming
field carries a finite amount of power, and the associated complex
travelling waves at the output ports carry a finite amount of
power: ${\bf T}$ is therefore bounded. More importantly, a real
system can only transmit a limited amount of information, because
a finite number of primary antennas is used: ${\bf T}$ is
therefore Hilbert-Schmidt. It follows that the integral operator
that maps the incoming field distribution onto the output ports
can be written in the form \cite{M}
\begin{equation}
\label{D_1} z_{p} = \int_{\cal A} \sum_{i} \sigma_{i} \, {\bf
U}_{i}^p {\bf V}_{i}^{\ast} (\hat{\boldsymbol \Omega}) \cdot {\bf
E}(\hat{\boldsymbol \Omega}) \, d {\boldsymbol \Omega} \mbox{.}
\end{equation}

It is a feature of the Hilbert-Schmidt decomposition that
$\{ {\bf V}_{i} (\hat{\bf \Omega}) : i \in 1,\cdots,P\}$ 
and $ \{ {\bf U}_{i}^{p} : p \in 1,\cdots,P \, ; \, i \in 1, \cdots,P 
\}$ are orthogonal
sets and therefore, according to (\ref{D_1}), the operation of a phased 
array can be regarded as first mapping the incoming field onto the 
beams, ${\bf V}_{i} (\hat{\bf \Omega})$, scaling by the singular 
values, $\sigma_{i}$, and reconstructing the complex travelling 
wave amplitudes at the output ports through the basis vectors 
${\bf U}_{i}^{p}$. Those reception patterns, ${\bf V}_{i} (\hat{\bf \Omega})$, 
associated with non-zero singular values span the field distributions at 
the input to which the phased array is sensitive, and those output vectors
${\bf U}_{i}^{p}$, associated with non-zero singular values span the 
vectors at the output to which the phased array can couple. These 
are the natural modes of a phased array, and we shall call them the 
{\em eigenfields}, by analogy with bolometric interferometers \cite{N}. 
In the case of fields that have been discretised for numerical calculation,
${\bf V}_{i} (\hat{\bf \Omega})$, ${\bf U}_{i}^{p}$, and $\sigma_{i}$
correspond to the singular vectors and singular values of the singular
value decomposition of the phased array's transmission matrix ${\bf T}$.

It can be shown, \cite{N}, although we shall not do so here, that 
the eigenfields on the sky are the eigenfunctions of the
Hermitian operator that maps the sky field onto the travelling
waves at the output ports, and then back again onto the sky in a
time reversed manner. The input eigenfields are those field
distributions that remain unchanged in form after this complete
round trip. Likewise, the output eigenfields are the eigenvectors
of the Hermitian operator that maps the travelling waves at the
output ports onto the sky in a time reversed manner, and then
forward again to the output ports. The output eigenfields are
those discrete vectors that can make this round trip unchanged in
form.

The eigenfields have many unique properties. First, they
represent the primary paths by which information can pass from the
sky to the output ports. The input eigenfields on the sky are
mutually orthogonal, even though the synthesised beams may be
non-orthogonal or linearly dependent. The output eigenfields,
which are actually discrete vectors in this case, are also
mutually orthogonal. Secondly, the number of singular values,
$\sigma_{i}$, greater than some threshold is the modal throughput
of the system. Thirdly, if two phased arrays are placed
side-by-side, with the intention of creating an aperture synthesis
interferometer, the input eigenfields having non-zero singular values
associated with the different telescopes are mutually orthogonal\cite{N},
which makes them ideal for analysing the behaviour of interferometers.

Now consider how the concepts of frames and eigenfields can be
used to construct powerful models of imaging phased arrays. The 
correlations at the output ports can be found through 
$Z_{pp'}=\langle z_{p} z_{p'}^{\ast} \rangle$; using (\ref{D_1}), 
\begin{equation} \label{E_1} Z_{pp'} = \sum_{i} \sum_{i'}
\sigma_{i} \sigma_{i'} {\bf U}_{i}^{p} {\bf U}_{i'}^{p' \ast}
\int_{\cal A} \int_{\cal A} {\bf V}_{i}^{\ast} (\hat{\boldsymbol
\Omega}_{1}) \cdot \overline{\overline{E}}(\hat{\bf
\Omega}_{1},\hat{\boldsymbol \Omega}_{2}) \cdot {\bf V}_{i'} 
(\hat{\boldsymbol \Omega}_{2}) \, d {\boldsymbol \Omega}_{1} \, 
d {\boldsymbol \Omega}_{2} \mbox{,}
\end{equation}
which is equivalent to (\ref{A_13}), but with the array described
in terms of its natural modes. (\ref{E_1}) describes the
projection of the coherence tensor of the source onto the input
eigenfields, the eigenfield coefficients are then multiplied  by
the singular values, and the correlations at the output ports
assembled through the output eigenfields. It is interesting to
observe that if a phased array is illuminated by an incoherent,
unpolarized, spatially uniform source, then using (\ref{A_11b})
with a uniform temperature $T_{s}$, and (\ref{E_1}), gives
\begin{equation} \label{E_2} Z_{pp'} = \frac{h \nu}{\exp [ h \nu /
kT] - 1} \sum_{i}  \sigma_{i}^{2}
{\bf U}_{i}^{p}{\bf U}_{i}^{p' \ast} \mbox{,}
\end{equation}
where we have followed a procedure similar to that described by  
(\ref{A_14c}). (\ref{E_2}) shows that the correlations at the output 
are merely those associated with the incoherent excitation of the output 
eigenfields. Indeed the total power, $W$, becomes
\begin{equation} \label{E_2b} 
W = \sum_{p} Z_{pp} = \frac{h \nu}{\exp [ h \nu /
kT] - 1}  \sum_{i}  \sigma_{i}^{2} \mbox{,}
\end{equation}
where we have used the orthogonality of the output eigenfields. 
Comparing with (\ref{A_14c}), 
\begin{equation}
\label{E_2c}
\sum_{p} A_{p} \, \Omega_{p} / 
\lambda^{2} = \sum_{p}  \sigma_{p}^{2} \, \leq \, M
\mbox{.}
\end{equation}
The total throughput is determined by the number of singular values
significantly greater than some threshold, usually determined by the 
noise. (\ref{E_2c}) can be regarded as a statement about the modal
throughput of a complete array; an important performance metric of 
phased arrays can therefore be determined from the synthesised beams 
alone.

\section{Noise}

Consider the case where the primary antennas and beam forming
network are are made up of passive components, such as a planar
array and microstrip coupling network. What noise sources appear 
at the output as a consequence of the losses? Because the signal 
is uncorrelated with the internally generated noise, the
correlation matrix at the output can be written ${\bf Z}' =
{\bf Z} + {\bf Z}_{N}$, where ${\bf Z}'$ is the combined output,
${\bf Z}$ the output due to the signal, and ${\bf Z}_{N}$  
the output due to the thermal noise from the losses.

If the array is illuminated by a uniform, unpolarized thermal 
source, and the losses in the array are at the same temperature as 
the source, $T_{s}$, the travelling waves at the output will be 
uncorrelated, and appear to originate from a thermal source having 
temperature $T_{s}$. Under these circumstances, using (\ref{A_14})
\begin{equation}
\label{E_5b} 
{\bf Z}' = \frac{h \nu}{\exp [ h \nu / kT_{s}] - 1} {\bf I}_{P} 
= \frac{h \nu^{3}}{c^{2}} \frac{1}{\exp [ h \nu / kT_{s}] - 1} 
\, {\bf R} + {\bf Z}_{N}
\mbox{,}
\end{equation}
where ${\bf R}$ has elements
\begin{equation}
R_{pp'} =  \int_{\cal A} \,  {\bf t}_{p}^{\ast} 
(\hat{\boldsymbol \Omega}) \cdot {\bf t}_{p'} 
(\hat{\boldsymbol \Omega}) \, d {\boldsymbol \Omega} 
\mbox{.}
\end{equation}
Rearranging,
\begin{equation}
\label{E_5c} 
{\bf Z}_{N} = \frac{h \nu}{\exp [ h \nu / kT_{s}] - 1} \left[
{\bf I}_{P} - \frac{1}{\lambda^{2}} \, {\bf R} \right]
\mbox{.}
\end{equation}
If the synthesised beam patterns are known, and all of the overlap 
integrals calculated, giving ${\bf R}$, the thermal noise and the 
correlations at the output, ${\bf Z}_{N}$, can be found.

Now consider the case where the phased array is not passive, for 
example when HEMT amplifiers are used, but the noise temperatures
of the primary receiving antennas are known, and equal. By definition, 
the noise temperature of a receiver is the temperature that a matched 
source would need to have in order to generate the same output
as a noiseless, but otherwise identical system. Let us call the 
associated noise correlation matrix at the output ${\bf Z}_{N}'$, where
${\bf Z}' = {\bf Z} + {\bf Z}_{N}'$, then to find ${\bf Z}_{N}'$ we 
simply need to calculate the coefficients of a uniform sky having the 
same physical temperature as the noise temperatures of the receivers. 
Using (\ref{A_11b}) and (\ref{A_13}) we get
\begin{equation} \label{E_4}  
{\bf Z}_{N} = \frac{h \nu}{\exp [ h \nu / kT_{N}] - 1} 
\, \frac{1}{\lambda^{2}} {\bf R} \mbox{,}
\end{equation}
where $T_{N}$ is the noise temperature of the individual receivers. 
In both cases, if the beams are orthogonal, the noise sources are 
uncorrelated. In the case where the primary receiver temperatures
are different, a similar scheme can easily be established.

\section{Fluctuations and Sensitivity}

The last step in the analysis is to determine the fluctuations in
power and the correlations between the fluctuations in power at
the output ports of a phased array once ${\bf Z}$, or ${\bf Z}'$,
is known; after all, it is the fluctuations that ultimately determine 
the sensitivity of an instrument. Following recent work on multimode 
detectors\cite{N,O}, it is straightforward to determine the 
fluctuations in any power-related measurement that is made at 
the output ports.

If ${\bf Z}$ is the coherence matrix of the complex travelling wave 
amplitudes at the output ports, and ${\bf W} \in {\mathbb C}^{P \times P}$ 
is a matrix that characterises the nature of the measurement being made, 
the expectation value of the measurement, $\langle P \rangle$, is given 
by
\begin{equation} \label{E_6}
\bar{P} = \langle P \rangle  =  \Delta \nu \, \mbox{Tr} {\bf W} {\bf Z} 
\mbox{,} 
\end{equation}
where $\Delta \nu$ is the RF bandwidth. (\ref{E_6}) can be appreciated by 
remembering that the trace of the product of two matrices constitutes an 
inner product in the abstract vector space of matrices. Thus, (\ref{E_6}) 
describes the way in which the state of coherence of the source projects 
onto the state of coherence to which the measurement system is sensitive. 
The structure of (\ref{E_6}) has be explored in detail, and leads to the 
coupled-mode theory of power detection\cite{O,P,Q}.

More significantly, it has also been shown\cite{P} that if two measurements
are made, one represented by ${\bf W}^{a}$ and one by ${\bf W}^{b}$, 
then the covariance of the measurements, $\mbox{Cov} \left[ P^{a},P^{b} 
\right] = \langle (P^{a}-\bar{P}^{a})(P^{b}-\bar{P}^{b}) \rangle$, is 
given by 
\begin{equation} \label{E_7}
\mbox{Cov} \left[ P^{a},P^{b} \right] = 
\frac{\Delta \nu}{\tau} \left[ \mbox{Tr} {\bf W}^{a} {\bf Z} {\bf W}^{b} 
{\bf Z} + \delta_{ab} h \nu_{0} \mbox{Tr} {\bf W}^{a} {\bf Z} \right]
\mbox{,}
\end{equation}
where $\nu_{0}$ is the central RF frequency, and $\tau$ the time for
which the output is integrated. (\ref{E_7}) is valid for thermal sources, and 
when the integration time is much longer than either the coherence time of the 
source or the intrinsic response time of the detector. It is valid, therefore,
for astronomical phased arrays. The first term in  (\ref{E_7}) can be
identified with the noise associated with classical waves, whereas the second
term with the noise associated with photon counting. Thus, (\ref{E_7})
contains the transition from fully bunched to Poisson statistics as the 
wavelength of operation moves from the microwave range through into the far 
infrared. (\ref{E_7}) contains a considerable amount of detail, including
partition noise, and the randomization of photon arrival times due to 
losses and imperfect quantum efficiencies.

We can imagine three principle measurements at the output ports. First 
consider a measurement having the matrix elements
\begin{equation} \label{E_8}
W_{ij}^{a} = \eta^{a} \delta_{ij} \delta_{ia} 
\mbox{,}
\end{equation}
where $i,j \in \{ 1, \cdots, P \}$, which is diagonal with just one 
non-zero element. Substituting (\ref{E_8}) in (\ref{E_6}) gives
\begin{eqnarray} \label{E_9}
\langle P^{a} \rangle  & = &  \Delta \nu \, \sum_{ij} W_{ij}^{a} Z_{ji} \\ \nonumber
 & = &  \Delta \nu  \, \eta^{a} \, \sum_{ij}  \delta_{ij} \delta_{ia} Z_{ji} \\ \nonumber
 & = &  \Delta \nu \, \eta^{a} Z_{aa}
\mbox{.} 
\end{eqnarray}
(\ref{E_8}) therefore corresponds to measuring the expectation value of the 
power at port $a$, using a detector having quantum efficiency $\eta^{a}$. 
Likewise, a power measurement at port $b$ is characterised by
\begin{equation} \label{E_10}
W_{ij}^{b} = \eta^{b} \delta_{ij} \delta_{ib} 
\mbox{.}
\end{equation}

In the case of a phased array where only powers are measured, 
which would correspond to an ordinary imaging array with the 
appropriate synthesised beams, (\ref{E_8}) and (\ref{E_10}) can
be substituted into (\ref{E_7}) to give
\begin{equation} \label{E_11}
\mbox{Cov} \left[ P^{a},P^{b} \right] = \frac{\Delta \nu}{\tau} 
\left[ \eta^{a} \eta^{b} |Z_{ab}|^{2} + \delta_{ab}  h \nu_{0}
\eta^{a} Z_{aa} \right] 
\mbox{.}
\end{equation} 

When we are interested in the true fluctuations in the power
at port $a$, as distinct from the fluctuations in the 
absorbed power, then $a=b$ and $\eta^{a}=\eta^{b}=1$, and the 
root mean square fluctuation, $\Delta P^{a}$, is given by
\begin{equation} \label{E_12}
\Delta P^{a} = \left\{ \mbox{Cov} \left[ P^{a},P^{a} \right] \right\}^{1/2} 
=  \left( \frac{\Delta \nu}{\tau} \right)^{1/2} \left[ 
|Z_{aa}|^{2} + h \nu_{0} Z_{aa} \right]^{1/2} 
\mbox{.}
\end{equation}

We can also form the quantity, which is the noise in the measurement 
normalised by the signal,
\begin{equation} \label{E_13}
\frac{\Delta P^{a}}{\langle P^{a} \rangle} = 
\frac{1}{\tau^{1/2}} \left[ \frac{1}{\Delta \nu}
+ \frac{h \nu_{0}}{\Delta \nu Z_{aa}} \right]^{1/2}
\mbox{.}
\end{equation}
or
\begin{equation} \label{E_14}
\frac{\Delta P^{a}}{\langle P^{a} \rangle} = 
\frac{1}{(\Delta \nu \, \tau)^{1/2}} \left[ 1 + 
\frac{\Delta \nu}{\bar{n}} \right]^{1/2}
\mbox{,}
\end{equation}
where $\bar{n} =  \Delta \nu Z_{aa} / h \nu_{0}$ is
the average photon rate. In the case where the photon rate is
much greater than the bandwidth, or equivalently many
photons arrive in the radiation's coherence time, the sensitivity
scales according to the radiometer equation; in the case where 
very few photons arrive in the coherence time, the sensitivity
scales according to photon counting.

Likewise the root mean square correlation, $\Delta P^{ab}$, between 
the fluctuations in two different power measurements, $a \neq b$, is 
given by (\ref{E_11}) as
\begin{equation} \label{E_15}
\Delta P^{ab} =  \left\{ \mbox{Cov} \left[ P^{a},P^{b} \right] 
\right\}^{1/2} =  \left( \frac{\Delta \nu}{\tau} \right)^{1/2} 
|Z_{ab}| 
\mbox{,}
\end{equation} 
and we also have
\begin{equation} \label{E_16}
\frac{\Delta P^{ab}}{\langle P^{a} \rangle^{1/2}\langle P^{b}\rangle^{1/2}} 
= \frac{1}{(\Delta \nu \, \tau)^{1/2}}  \frac{|Z_{ab}|}{
(Z_{aa}Z_{bb})^{1/2}} = \frac{1}{(\Delta \nu \, \tau)^{1/2}} |\Gamma^{ab}|
\mbox{,}
\end{equation}
where $|\Gamma^{ab}|$ is the correlation coefficient. (\ref{E_16})
is the Hanbury Brown-Twiss effect for phased arrays. It does not show Poisson
noise because the Poisson term corresponds to photon arrival times that are
uncorrelated. In the case where the synthesised beams are orthogonal, 
and the source uniform, $Z_{ab} = 0 $, and the fluctuations in the powers 
measured at different ports are uncorrelated. In the case of non-orthogonal 
beams, or a non-uniform source on the sky, the fluctuations are correlated, 
and the correlations must be taken into account when calculating the noise in 
the reconstructed image. 

Now consider the case where we wish to measure the correlations between
the travelling waves at the output ports, say $a$ and $b$, for the purpose of 
recovering an image, as in done in interferometry. In this context, there are 
two measurement-system matrices of interest:
\begin{eqnarray} \label{E_17}
R_{ij}^{ab} & = & \eta^{ab} \frac{1}{2} ( \delta_{ia} \delta_{jb} + 
\delta_{ib} \delta_{ja} ) \\ \nonumber
I_{ij}^{ab} & = & \eta^{ab} \frac{i}{2} ( \delta_{ia} \delta_{jb} - 
\delta_{ib} \delta_{ja} )
\mbox{,}
\end{eqnarray}
where $a \neq b$, and the $i$ in the prefactor corresponds to the unit 
imaginary, rather than the index, with no confusion. The two 
matrices comprising the elements listed in (\ref{E_17}) are Hermitian,
and characterise realizable measurements.

Substituting (\ref{E_17}) in (\ref{E_6}),
\begin{eqnarray} \label{E_18}
\langle R^{ab} \rangle & = & \Delta \nu \, \eta^{ab} \, \mbox{Re} 
Z_{ab}\\ \nonumber
\langle I^{ab} \rangle & = & \Delta \nu \, \eta^{ab} \, \mbox{Im} 
Z_{ab}
\mbox{,}
\end{eqnarray}
and the two measurements correspond to finding the real and 
imaginary parts of the correlations between ports $a$ and $b$,
which is equivalent to finding the in and out of phase components
of the fringe formed when the signals at $a$ and $b$ are 
combined. More specifically, the matrices correspond to using 
an analogue correlator to measure the $\cos$ and $\sin$ fringes.

We can now substitute (\ref{E_17}) into (\ref{E_7}) to give
\begin{equation} \label{E_19}
\mbox{Cov} \left[ R^{ab},R^{ab} \right] = \frac{\Delta \nu}{\tau} 
\left[ (\eta^{ab})^{2} \frac{1}{2} ( Z_{aa}Z_{bb} + \mbox{Re} Z_{ab}^{2} ) 
+ h \nu_{0} \eta^{ab} \mbox{Re} Z_{ab} \right]
\mbox{,}
\end{equation}
giving a noise to signal ratio of
\begin{equation} \label{E_20}
\frac{\Delta R^{ab}}{\langle R^{ab} \rangle} = 
\frac{1}{(\Delta \nu \, \tau)^{1/2}} \frac{
\left[  \frac{1}{2} (Z_{aa}Z_{bb} + \mbox{Re} Z_{ab}^{2} ) 
+ h \nu_{0} \mbox{Re} Z_{ab} \right]^{1/2}}{\mbox{Re} Z_{ab}}
\mbox{,}
\end{equation}
for $\eta^{ab}=1$, which corresponds to the actual power.
For a completely incoherent source, and orthogonal beams,
$Z_{ab} = 0$, the noise becomes
\begin{equation} \label{E_21}
\Delta R^{ab} =  \frac{1}{(\Delta \nu \, \tau)^{1/2}} \frac{1}{\sqrt{2}}
\left( \langle P_{aa} \rangle \langle P_{bb} \rangle \right)^{1/2}
\end{equation}
and the noise is classical, with the fluctuations being proportional to
the geometric means of the two power levels, which is a well known feature
of noise in interferometers.

In the case where the two signals at $a$ and $b$ are fully coherent, 
$Z_{aa}Z_{bb} = |Z_{ab}|^{2}$, we get
\begin{equation} \label{E_22}
\frac{\Delta R^{ab}}{\langle R^{ab} \rangle} = 
\frac{1}{\tau^{1/2}} \left[ \frac{1}{\Delta \omega}
+ \frac{ h \nu_{0}}{\Delta \omega |Z_{ab}|
\cos \theta_{ab} } \right]^{1/2}
\mbox{,}
\end{equation}
or
\begin{equation} 
\label{E_23}
\frac{\Delta R^{ab}}{\langle R^{ab} \rangle} = 
\frac{1}{(\Delta \nu \, \tau)^{1/2}} \left[ 1
+ \frac{\Delta \nu}{\bar{n}_{c}} \right]^{1/2}
\mbox{,}
\end{equation}
where $\bar{n}_{c}$ is the mean number of photons in the in-phase
correlated power, and $\theta_{ab}$ is the phase of $Z_{ab}$.
Thus, for weak fringes, and even in the nulled region
of a strong fringe, classical noise is swapped for Poisson noise. In 
general, the nature of the noise will change throughout the fringe.

Similarly, we have
\begin{equation} \label{E_24}
\mbox{Cov} \left[ I^{ab},I^{ab} \right] = \frac{\Delta \nu}{\tau} 
\left[ (\eta^{ab})^{2} \frac{1}{2} ( Z_{aa}Z_{bb} - \mbox{Re} Z_{ab}^{2} ) 
+ h \nu_{0} \eta^{ab} \mbox{Im} Z_{ab} \right]
\mbox{,}
\end{equation}
giving
\begin{equation} \label{E_25}
\frac{\Delta I^{ab}}{\langle I^{ab} \rangle} = 
\frac{1}{(\Delta \nu \, \tau)^{1/2}} \frac{
\left[  \frac{1}{2} (Z_{aa}Z_{bb} - \mbox{Re} Z_{ab}^{2} ) 
+ h \nu_{0} \mbox{Im} Z_{ab} \right]^{1/2}}{\mbox{Im} Z_{ab}}
\mbox{,}
\end{equation}
for $\eta^{ab}=1$.

For incoherent travelling waves, the noise is the same as for the
in-phase measurement, (\ref{E_21}). For fully coherent signals,
\begin{equation} 
\label{E_26}
\frac{\Delta I^{ab}}{\langle I^{ab} \rangle} = 
\frac{1}{\tau^{1/2}} \left[ \frac{1}{\Delta \nu}
+ \frac{h \nu_{0}}{\Delta \nu |Z_{ab}|
\sin \theta_{ab}} \right]^{1/2}
\mbox{.}
\end{equation}
or
\begin{equation} 
\label{E_27}
\frac{\Delta I^{ab}}{\langle I^{ab} \rangle} = 
\frac{1}{(\Delta \nu \, \tau)^{1/2}} \left[ 1
+ \frac{\Delta \nu}{\bar{n}_{s}} \right]^{1/2}
\mbox{,}
\end{equation}
where $\bar{n}_{s}$ is the mean number of photons in the quadrature
component of the correlated power. Overall, the same behaviour is seen 
as for the in-phase component.

Finally, we can correlate the fluctuations in the in-phase and quadrature
measurements, giving
\begin{equation} \label{E_28}
\mbox{Cov} \left[ R^{ab},I^{ab} \right] = \frac{\Delta \nu}{\tau} 
\left[ (\eta^{ab})^{2} \frac{1}{2} \mbox{Im} Z_{ab}^{2} \right] 
\mbox{,}
\end{equation}
and
\begin{equation} \label{E_29}
\frac{\Delta RI^{ab}}{\langle R^{ab} \rangle^{1/2} 
\langle I^{ab} \rangle^{1/2}} = 
\frac{1}{(\Delta \nu \, \tau)^{1/2}}  \frac{1}{\sqrt{2}}
\left[  \frac{\mbox{Im} Z_{ab}^{2}} {\mbox{Re} Z_{ab} \, \mbox{Im} Z_{ab}}
\right]^{1/2}
\mbox{,}
\end{equation}
giving
\begin{equation} 
 \label{E_30}
\frac{\Delta RI^{ab}}{\langle R^{ab} \rangle^{1/2} 
\langle I^{ab} \rangle^{1/2}} = 
\frac{1}{(\Delta \nu \, \tau)^{1/2}} 
\mbox{.}
\end{equation}
It is also possible to calculate the correlations between the fluctuations 
in the real and imaginary parts of fringe measurements on two different 
pairs of ports. 

\section{Conclusion}

We have analyzed the behaviour of phased arrays from a functional
perspective, and shown that their operation is intimately
related to the mathematical theory of frames. In cases where the beams
are non-orthogonal, or even linearly dependent, image reconstruction 
can be carried out using dual beams. The theory of frames 
allows one to assess, in a straightforward manner, whether the 
powers or correlations at the output ports of a phased array contain sufficient 
information to allow some class of field or intensity distribution to be 
reconstructed precisely. We have also identified the natural modes of phased 
arrays, which are important for understanding information throughput and 
aperture synthesis interferometry.

In order to calculate the behaviour of an imaging phased array it is only 
necessary to know the synthesised reception patterns, which may be 
non-orthogonal. It is not necessary to know anything about the internal 
construction of the array itself. As a consequence, data can be taken from 
experimental measurements or from electromagnetic simulations. The ability 
to assess behaviour simply from the synthesised beams separates the 
process of choosing the best beams for a given application from the process 
of realizing the beams in practice. It also suggests important techniques for 
analysing experimental data.

Our model allows the straightforward calculation of quantities
such as the correlations in the fluctuations at the output ports of an 
phased array. Indeed, sources can be constructed through expressions
of the kind (\ref{A_12}), the correlations at the output ports calculated 
through (\ref{A_13}), or in the modal case (\ref{E_1}), system noise can 
be included through (\ref{E_5c}) and (\ref{E_4}), 
and the expectation values, fluctuations, and correlations between fluctuations 
in measurements at the output determined through (\ref{E_6}) and (\ref{E_7}). 
The whole procedure only requires simple matrix algebra. The scheme is 
conceptually and numerically powerful\cite{R}, and in an upcoming paper 
we shall present simulations showing dual beams, natural modes, and image 
reconstructions.

\end{document}